\documentclass[sigconf]{acmart}
\usepackage{bm}

\AtBeginDocument{%
  }

\copyrightyear{2024}
\acmYear{2024}
\setcopyright{licensedusgovmixed}\acmConference[GeoAnomalies'24]{1st ACM SIGSPATIAL International Workshop on Geospatial Anomaly Detection}{October 29, 2024}{Atlanta, GA, USA}
\acmBooktitle{1st ACM SIGSPATIAL International Workshop on Geospatial Anomaly Detection (GeoAnomalies'24), October 29, 2024, Atlanta, GA, USA}
\acmDOI{10.1145/3681765.3698449}
\acmISBN{979-8-4007-1144-2/24/10}

\begin{document}

\title{Spatiotemporal Modeling and Forecasting at Scale with Dynamic Generalized Linear Models}

\author{Pranay Pherwani}
\authornote{Both authors contributed equally to this research.}
\affiliation{%
  \institution{STR}
  \city{Woburn}
  \state{MA}
  \country{USA}
}
\email{pranay.pherwani@str.us}

\author{Nicholas Hass}
\authornotemark[1]
\affiliation{%
  \institution{STR}
  \city{Woburn}
  \state{MA}
  \country{USA}
}
\email{nicholas.hass@str.us}

\author{Anna K. Yanchenko}
\affiliation{%
  \institution{STR}
  \city{Woburn}
  \state{MA}
  \country{USA}
}
\email{anna.yanchenko@str.us}

\renewcommand{\shortauthors}{Pherwani, Hass, Yanchenko}

\begin{abstract}
 Spatiotemporal data consisting of timestamps, GPS coordinates, and IDs occurs in many settings.  Modeling approaches for this type of data must address challenges in terms of sensor noise, uneven sampling rates, and non-persistent IDs.  In this work, we characterize and forecast human mobility at scale with dynamic generalized linear models (DGLMs).  We represent mobility data as occupancy counts of spatial cells over time and use DGLMs to model the occupancy counts for each spatial cell in an area of interest. DGLMs are flexible to varying numbers of occupancy counts across spatial cells, are dynamic, and easily incorporate daily and weekly seasonality in the aggregate-level behavior.  Our overall approach is robust to various types of noise and scales linearly in the number of spatial cells, time bins, and agents.  Our results show that DGLMs provide accurate occupancy count forecasts over a variety of spatial resolutions and forecast horizons.  We also present scaling results for spatiotemporal data consisting of hundreds of millions of observations.  Our approach is flexible to support several downstream applications, including characterizing human mobility, forecasting occupancy counts, and anomaly detection for aggregate-level behaviors.
\end{abstract}




\maketitle

\section{Introduction}

Spatiotemporal data occurs in a variety of different settings such as traffic management \cite{Yuan:2021, Tebaldi2002}, ecology \cite{HEFLEY2017206}, environmental monitoring \cite{Amato:2020vb} and epidemic control \cite{10.3389/fpubh.2021.641253}.  There are several characteristics of GPS trajectory spatiotemporal data that make accurate modeling of this data challenging, including sensor noise and uneven sampling rates of observations.  Additionally, spatiotemporal data often exhibits heterogeneous dependencies over space, time, and between agents or entities in the trajectory data, and accurately capturing these dependencies while modeling spatiotemporal data at scale is a significant challenge.  In this work, we propose an efficient, flexible, and scalable approach to model and forecast spatiotemporal data at scale using dynamic generalized linear models (DGLMs) \cite{West1985a, West1997}.

Our approach focuses on modeling human mobility data specifically and we are interested in modeling mobility data for hundreds of thousands of agents for months of time over areas of thousands of km$^2$.  Our setting requires models that accurately capture normal human mobility to support a variety of downstream applications, including forecasting and anomaly detection.  We require modeling approaches that are efficient, probabilistic, robust to various types of noise, and can incorporate important dependencies exhibited in the data, such as daily seasonality in human mobility.  We choose DGLMs in this setting due to their ability to meet these requirements.  

\subsection{Related Work}

Accurate models for human mobility data support many downstream applications, including traffic flow forecasting \cite{BARBOSA20181, 10.5555/3060832.3060987, Tebaldi2002}, urban planning \cite{8822969}, point of interest demand analysis \cite{10.1145/3097983.3098168}, epidemic modeling \cite{8822969}, and anomaly detection.    Related work in mobility modeling can largely be separated into modeling fine-grained agent movement or larger scale, population-level movement \cite{BARBOSA20181} and can also be separated by which modes of transportation are modeled, e.g. pedestrian movement, vehicular movement, etc. \cite{8822969}.  
 
There are a variety of approaches to modeling spatiotemporal data generally and human mobility data specifically, ranging from traditional point process models to more recent deep learning approaches.  Mechanistic models, such as the gravity model, model mobility flow between locations as inversely related to the distance between those locations \cite{Zipf}.  Bayesian latent factor models were used to model demand patterns based on points of interest in \cite{10.1145/3097983.3098168}.  Gaussian processes are commonly used statistical models for spatiotemporal data; recent extensions include capturing social-network dependencies between agents in animal movement data \cite{Scharf:2018} and combining Gaussian processes with deep learning to predict origin-destination flows in taxi data \cite{Steentoft:2023vp}. Finally, deep learning approaches are becoming more popular for spatiotemporal data, for example, to model and predict city-wide traffic patterns with LSTMs \cite{10.5555/3060832.3060987}.


Dynamic generalized linear models are Bayesian state space models that are flexible to various observation distributions \cite{West1985a, West1997}. They have been successfully applied in a variety of settings involving large-scale data, including retail forecasting \cite{BerryWest2018DCMM, YanchenkoEtAl2021} and online advertising for modeling traffic flows through a website \cite{ChenBanksWest2018}.  Critically, in our setting observations are preprocessed into occupancy counts in space-time bins and DGLMs can flexibly model count time series data with a variety of sparsity and over-dispersion levels \cite{BerryWest2018DCMM}.  Related work using state space models for spatiotemporal data includes traffic flow prediction and modeling with Kalman filters \cite{EMAMI2020102025, 6580568}, though only for settings that admit normally distributed observations, modeling rainfall data and water temperature \cite{Stroud:2001} and modeling tornado reports and mortality ratios \cite{doi:10.1080/01621459.2015.1129968}. 

In this work, we focus on aggregate-level modeling of human mobility over long and short time horizons and our approach is applied to data with walking and driving movement patterns.  While our results focus on multi-step ahead forecasting of occupancy counts, we propose a modeling approach that is flexible to a variety of downstream applications.  To this end, we require models that are probabilistic to accurately quantify uncertainty, can be used for multi-step ahead forecasting, are able to flexibly incorporate external information (e.g. point of interest data, holiday effects, seasonality, etc.), and can accurately model count time series.  We need to meet these modeling goals with an approach that can scale to large amounts of data for city-wide human mobility modeling.  

Our approach meets all of these modeling requirements by preprocessing human mobility data into occupancy counts in space-time bins and then modeling and forecasting these counts with DGLMs. DGLMs additionally have very competitive forecast accuracy compared to competing modeling approaches  \cite{BerryWest2018DCMM}. Other related work does not meet all of these criteria; mechanistic and deep learning approaches often do not represent uncertainty and can be challenging to scale. Statistical models like Gaussian processes and extensions with complex dependencies can also be challenging to scale.  Previous state space modeling approaches have in general been applied at a smaller scale than in this work, or to different spatiotemporal data that is not human mobility data.  To the best of our knowledge, our application of DGLMs, and specifically the sparse count mixture variations \cite{BerryWest2018DCMM, YanchenkoEtAl2021}, to large scale mobility data modeling and forecasting is novel.

\subsection{Contributions}

In this work, we present a flexible and scalable approach to modeling human mobility data with dynamic generalized linear models.  We use a preprocessing approach to convert raw spatiotemporal data to occupancy counts in space-time bins.  Occupancy counts for each spatial bin are modeled with DGLMs and we present forecasting results for several spatial resolutions and forecast horizons; our forecast point accuracy does not degrade between fifteen-minute (1-step ahead) forecasts up to twenty-four hour (96-step ahead) forecast horizons. Across a variety of different levels and patterns of observed occupancy counts, we are able to accurately model and forecast with this common modeling approach.  Finally, we present details on the scaling and implementation of our approach, which is linear in the number of observations, spatial cells, and time bins.  While our results focus on human mobility data, this work is applicable to other application areas for spatiotemporal data where accurate probabilistic modeling at scale is the main interest.  Our approach supports fully online preprocessing, updating, and forecasting, and anomaly detection for aggregate behaviors.

\section{Data and Application Goals}
In this work, we focus on modeling trajectory data, where each observation consists of an agent ID, latitude, longitude, and timestamp.  Observations can be unevenly sampled in time and space, as there is often sensor noise in the latitude/longitude values and IDs for agents may not be persistent.  Additionally, human mobility data tends to be largely stationary, with movement occurring between stationary locations.  Aggregate occupancy counts exhibit a large range of values, from very sparsely occupied cells to densely populated locations.  Our modeling approach needs to address these challenges.  Our goal for this work is to flexibly model large-scale human mobility data to support a variety of downstream applications, such as forecasting traffic flow and performing aggregate-level anomaly detection.

The results presented in this work use simulated trajectory data, following the generative process described in \cite{Tsiligkaridis:2024}.  Specifically, the generative data model is parameterized by sequences of persistent locations that are associated with each agent in the synthetic data. The data generating process assumes that agents often: (1) have recurring behaviors in the same locations (e.g. home and work locations), (2) live and behave similarly to other agents, and (3) travel along roads between their persistent locations \cite{Tsiligkaridis:2024}.  

\begin{figure}[h]
  \centering
  \includegraphics[width=0.8\linewidth]{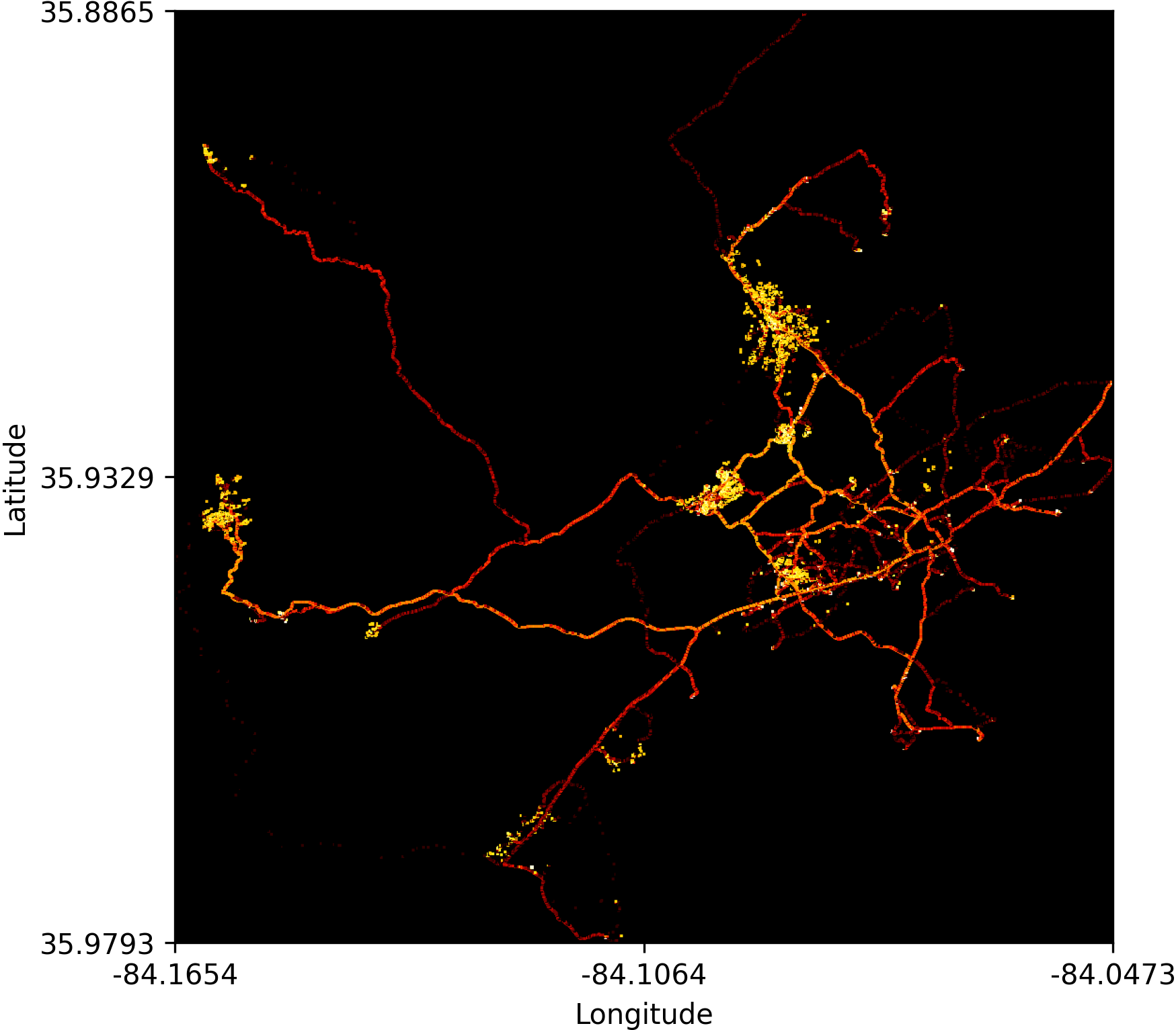}
  \caption{Synthetic human mobility data for 10909 agents over a 28-day time period.  Brighter colors indicate a higher density of observations.  The area of interest is large, resulting in both dense and sparse areas to model.}
  \label{fig:data}
\end{figure}

Persistent locations are randomly assigned to each agent using foundation data from \cite{USA-structures, PlanetSense, OSM}, and the assignment of persistent locations encourages similarity among groups of agents (e.g. agents share the same office location or have nearby houses).  Agents travel on roads between persistent locations; the synthetic data uses an Open Street Maps (OSM) road network from the OSMnx Python package \cite{Boeing:2017}, which results in a high frequency human mobility dataset.   For our experiments below, we start with this simulated trajectory data sampled every 60 seconds for 10909 agents over a 28-day period (Figure~\ref{fig:data}).  We do not assume specific knowledge of this data generating procedure in our results below. 

\begin{figure}[h]
  \centering
  \includegraphics[width=0.7\linewidth]{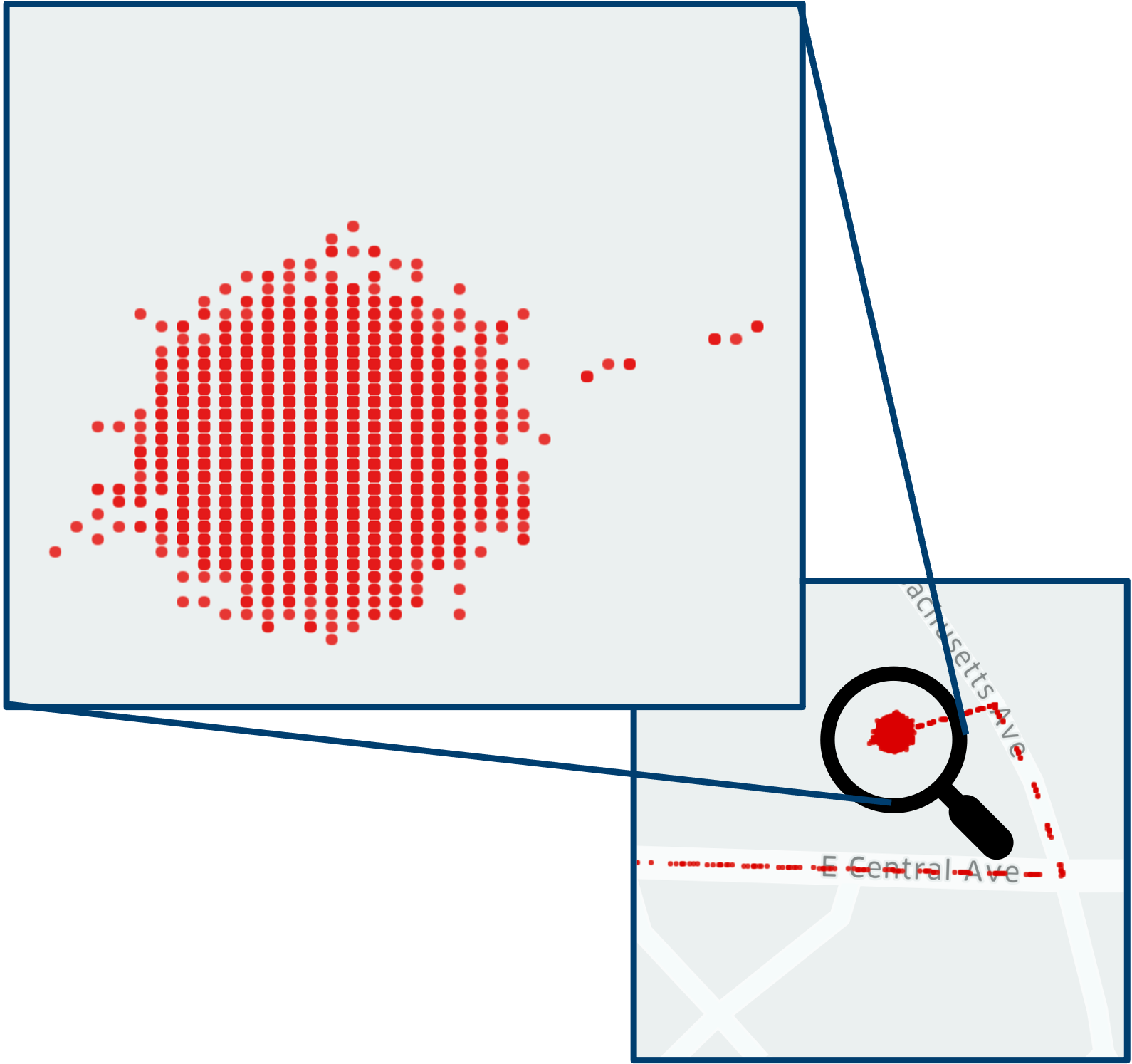}
  \caption{Simulated sensor noise from our synthetic human mobility data.  We encode our mobility data into space-time bins to be robust to sensor noise.}
\end{figure}

\section{Spatiotemporal Modeling with DGLMs}
\subsection{Data Preprocessing}

To meet our goal of flexibly modeling human mobility data at scale, we first preprocess the raw trajectory data  (ID, latitude, longitude, timestamp observations) into occupancy counts of unique agents in space-time bins.  This discretization of space and time and reduction from individual agent-level trajectories to aggregate occupancy counts significantly reduces the volume of data and enables parallelization across relevant dimensions in the modeling.  Additionally, converting the data to occupancy counts in discretized space-time bins provides robustness to sensor noise, uneven sampling rates of the data over time, and ID dropout or confusion. While discretizing the data over space and time and then aggregating into occupancy counts loses information at the individual observation level, this approach supports our goal of flexibly modeling at scale.

\begin{figure*}[h]
  \centering
  \includegraphics[width=0.9\textwidth]{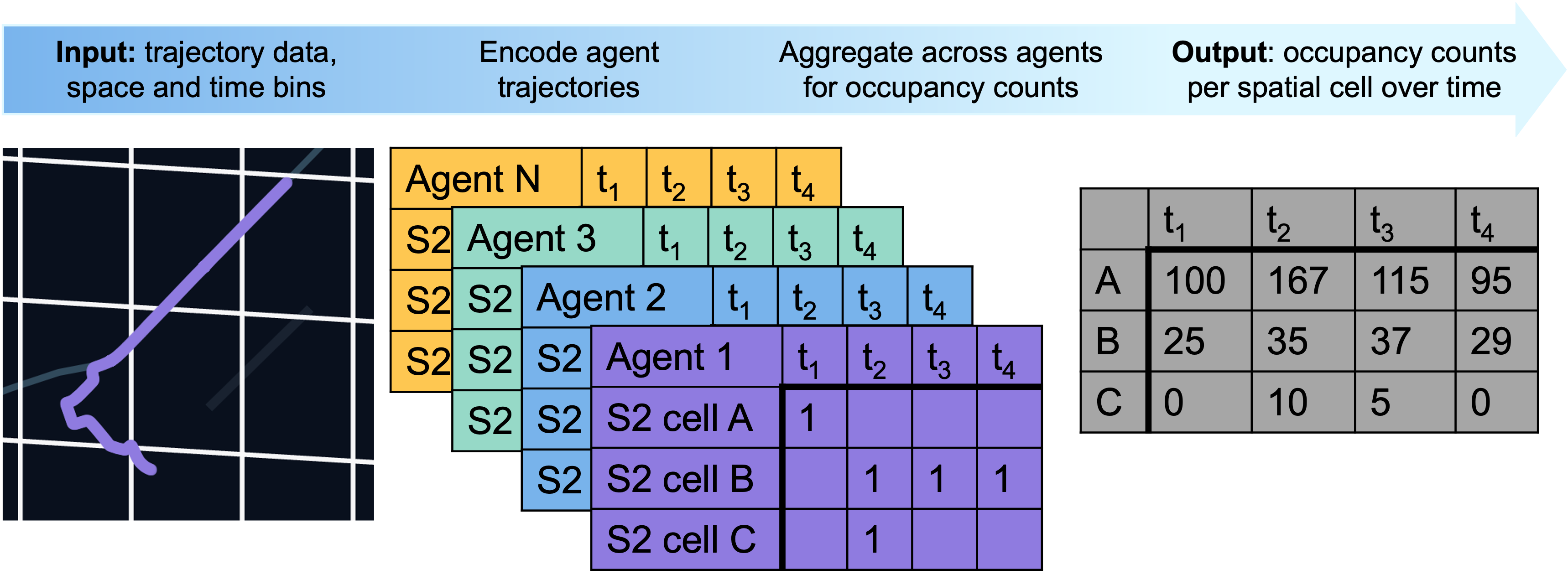}
  \caption{Preprocessing procedure to convert raw mobility data into occupancy counts in space-time bins.  The encoding occurs in parallel across all observations, making this approach highly scalable.}
   \label{fig:preprocessing}
\end{figure*}

Our preprocessing approach is outlined in Figure~\ref{fig:preprocessing}.  First, we encode all latitude, longitude values into a spatial bin.  In this work, we use S2 cells \cite{S2} to discretize space, though our approach is flexible to any choice of geohashing algorithm, including geohash \cite{geohash} or H3 cells \cite{H3}.  Next, we convert the raw timestamp values into temporal bins based on a chosen temporal resolution.  This encoding of space and time can occur in parallel across observations and does not depend on the choice of spatial or temporal resolution (unless the geohashing algorithm takes longer to encode for finer resolutions).  Finally, to form the occupancy counts, we count the number of unique agents in each space-time bin.  We allow each agent to be in multiple spatial bins within the same time bin, though this assumption is not required; the first, last, mean, etc. location for each time bin could be selected instead.  This procedure converts the raw trajectory observations into aggregate occupancy counts over spatial and temporal bins for the entire dataset of interest in a scalable manner. 

\begin{figure*}[h]
  \centering
  \includegraphics[width=0.9\textwidth]{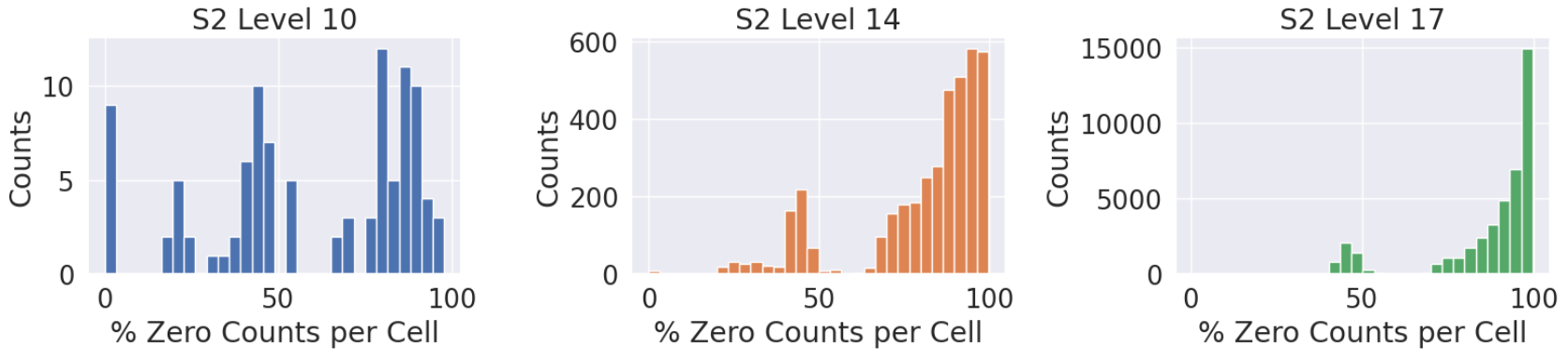}
  \caption{Sparsity (or percent of zero occupancy counts per spatial cell) for three different spatial resolutions.  At fine spatial resolutions (e.g. S2 cell level 17), the majority of spatial cells have zero occupancy for more than 50\% of the 15-min time bins over the month of our synthetic data. }
   \label{fig:sparsity}
\end{figure*}

In this work, we consider three different spatial resolutions to demonstrate the flexibility of our approach; summary statistics for these spatial resolutions are given in Table~\ref{tab:s2}. The temporal resolution for our results is 15-minute time bins, for a total of 2688 time bins over 28 days of synthetic data.  Finer spatial and/or temporal resolutions decrease the loss of spatial information from discretizing the data, but increase the sparsity in the resulting occupancy counts and increase the sensitivity to noise.  The degree of sparsity for each spatial cell increases at finer spatial resolutions (Figure~\ref{fig:sparsity}), which also impacts the modeling. In Table~\ref{tab:s2}, at finer spatial resolutions, more spatial cells require mixture models (DCMMs or DLMMs, described below) to model the observed occupancy counts.      

\subsection{Dynamic Generalized Linear Models}

After converting the spatiotemporal data to occupancy counts in space-time bins, we model the resulting data with  dynamic generalized linear models (DGLMs) \cite{West1985a, West1997}.  DGLMs are Bayesian state space models that can easily incorporate seasonality, covariates, and holiday effects.  DGLMs have straightforward probabilistic forecast distributions, and are flexible to a variety of occupancy count distributions, from very high-levels of counts in each spatial cell, to very sparse cells. In this work, we fit one DGLM to each spatial cell in our area of interest.

Let $y_t$ be the observed occupancy count at time bin $t$ for one spatial cell.  We assume that $y_t$ comes from an exponential family distribution and define a DGLM to model this sequence of occupancy counts as follows \cite{West1997}:

\begin{equation}\label{eq:dglm}
\begin{split}
  y_t &\sim ExpFamily(\eta_t) \\
  \lambda_t &= \mathbf{F}'_t\bm{\theta}_t = g_{link}(\eta_t) \\
  \bm{\theta}_t &= \mathbf{G}_t\bm{\theta}_{t-1} + \bm{\omega_t}, \\
  \bm{\omega}_t &\sim (\mathbf{0}, \bm{W}_t), 
  \end{split}
\end{equation}

\noindent where $g_{link}(\cdot)$ is the link function for the chosen exponential family form, $\eta_t$ is the natural parameter, $\bm{\theta}_t$ is the state vector, $\bm{F}_t$ is a vector of known dynamic covariates, $\bm{G}_t$ is a known state matrix, $\bm{\omega}_t$ is the evolution noise, and $\mathbb{E}(\bm{\omega}_t) = \bm{0}, \mathbb{V}(\bm{\omega}_t) = \bm{W}_t$.  Dynamic covariates, such as point of interest information, holiday effects, etc. are incorporated into the $\bm{F}_t$ vector, and seasonality is also incorporated via $\bm{F}_t$ and $\bm{G}_t$.  See Section~\ref{subsec:dglm-learning} and \cite{West1997} for more details about model inference.  

The  $\bm{W}_t$ matrices represent the evolution variance and are specified in terms of component discounting  \cite{West1997}.   $\bm{W}_t$ is partitioned into components for the trend, seasonal terms, and covariates, and each component includes a specific discount factor.  Discount factors specify the level of change in the stochastic state evolution and how quickly the DGLM state vectors adapt to new information; discount factors close to 1 indicate very little stochastic noise in the state evolution, while discount factors closer to 0 indicate large variation in the respective state vector, allowing the DGLMs to adapt very rapidly to incoming information and ``discount'' the value of older observations \cite{West1997}.  In practice, discount factors are rarely set lower than about 0.9.  Discount factors are the main hyper-parameters to tune for DGLMs and can be manually selected based on domain knowledge or tuned with a grid search.    

In this work, we automatically choose the specific DGLM form based on the observed level of occupancy counts in the initial training data. Due to the choice of conjugate priors for the parameters in the DGLM, these models have closed-form forecast distributions, which makes forecasting efficient \cite{West1997}.  Spatial cells with high levels of occupancy counts (defined as a mean occupancy count over the training period of > 50) are fit with a normal dynamic linear model (DLM), where the occupancy counts are assumed to follow a normal distribution and the forecasts are t-distributed.   For spatial cells with a mean occupancy count over the training data of < 50, we fit a Poisson DGLM; forecasts are Negative Binomial distributed.  We also use mixture model extensions to DGLMs to model and forecast the sparse occupancy counts that we observe in a variety of spatial cells. Dynamic count mixture models (DCMMs) \cite{BerryWest2018DCMM} are a mixture of a Bernoulli DGLM (to model and predict zero counts) and a Poisson DGLM (to model and predict the non-zero counts) and can model a variety of different observed sparsity levels. DCMMs also include a random effects extension for over-dispersed count data.  The forecast distribution for the DCMM is the product of Beta-Bernoulli distributed forecasts from the Bernoulli DGLM and Negative Binomial  forecasts from the Poisson DGLM.  We fit DCMMs to spatial cells that have a mean < 50 and an observed percent of zero occupancy counts that is greater than 15\%; as seen in Table~\ref{tab:s2}, we fit DCMMs to many spatial cells across all spatial resolutions, but especially for finer resolutions.  Additionally, for spatial cells that exhibit high average occupancy count levels, but still have a degree of sparsity > 15\%, we use dynamic linear mixture models (DLMMs) \cite{YanchenkoEtAl2021}, which are a mixture of a Bernoulli DGLM and a normal DLM; forecast distributions are  a mixture of samples from the Beta-Bernoulli and the t-distribution.

 We include a trend term (local level or dynamic intercept) and daily seasonality in the DGLMs for all spatial cells. We also need to specify trend and seasonality discount factors for all DGLMs, random effect discount factors for the DCMMs and Poisson DGLMs, and a discount factor on the stochastic variance of the DLM and DLMM observations.  Based on offline tuning experiments, we set the seasonality discount factor to 0.994 and the random effect discount factor or the stochastic variance discount factor to 0.9 for all spatial cells.  We tune the trend discount factor using a grid search over the possible values of $[0.96, 0.97, 1.0]$ for each spatial cell individually and select the value that minimizes the mean absolute error of the forecast median over the first 3 days of data (results below are presented for the following 25 days of simulated data).  Additional tuning of all discount factors via a grid search is straightforward, though does add to the computation time.

DGLM updates are linear in the number of timesteps and do not require expensive Markov-chain Monte Carlo or sequential Monte Carlo, which makes these models efficient for inference and forecasting at scale in our application.  Additionally, this updating occurs in an online, sequential manner, enabling our approach to support streaming data settings. Updating and forecasting only require the parameter values from the previous timestep, making the DGLMs efficient in terms of memory, as well.  We fit one model per spatial cell and these models are updated in parallel across space. 

\begin{table*}
  \caption{Spatial resolution and modeling details for different S2 cell levels; S2 cell information from \cite{S2}.  Finer spatial resolutions exhibit more sparsity, requiring the majority of spatial cells to be DCMMs or DLMMs, as compared with the coarser cells.}
  \label{tab:s2}
  \begin{tabular}{cccccccc}
    \toprule
    S2 Cell Resolution & Avg. Cell  & Avg. Cell  & \# Spatial Cells /  & \# DLMs & \# DLMMs & \# Poisson DGLMs & \# DCMMs \\
   & Area & Edge Length &  \# Models \\
    \midrule
    10 & 81.07 $\mbox{km}^2$ & 8.8 km & 103 & 9 & 30 & 0 & 64\\
    14 & 0.32 $\mbox{km}^2$ & 550 m &  3938 & 7 & 41 & 1 & 3889\\
    17 &  4948 $\mbox{m}^2$  &  70 m & 42038 & 0 & 35 & 1 & 42002 \\
  \bottomrule
\end{tabular}
\end{table*}

\subsection{Forecast Metrics}
We evaluate how well our modeling approach fits the observed occupancy counts in terms of both point and uncertainty forecast metrics.  We evaluate the accuracy of our point forecasts using root-mean squared error (RMSE), mean absolute error (MAE), and zero-adjusted percent error (ZAPE); ZAPE explicitly evaluates how well our models predict zero vs. non-zero occupancy counts.  The specific form of ZAPE used here follows \cite{YanchenkoEtAl2021} and the loss function is defined for a non-negative count $y$ and forecast $f$ as:
\begin{equation}
 \mathcal{L}_{ZAPE} = \bm{1}(y = 0)\dfrac{f}{(1+f)} + \bm{1}(y > 0)\dfrac{|y - f|}{y},
 \end{equation} 
 \noindent where $\bm{1}(\cdot)$ is the indicator function. The RMSE and MAE metrics are on the same scale as the data (occupancy counts) and ZAPE is a percent error metric. Point forecast metrics below are calculated using the forecast median. We also use coverage to evaluate the calibration of our forecast distributions; for well-calibrated uncertainty estimates, we expect our empirical coverage to be close to the theoretical coverage value. 
 
\subsection{Scaling Summary}
Our preprocessing approach to convert the raw trajectory data observations to occupancy counts is linear in the number of observations (or linear in the number of agents if each agent has approximately the same number of observations).  Modeling and forecasting with DGLMs is linear in the number of time bins, with 1 DGLM fit to each spatial cell. In summary, our overall approach is linear in the number of observations, time bins, and spatial bins.

\subsection{Extensions}

There are several straightforward extensions to our approach that would only require updates to the preprocessing procedure, but not to the overall modeling approach.  While we focus on modeling occupancy counts in this work, extending to the common setting of modeling transitions between spatial cells, or origin-destination matrices \cite{BARBOSA20181}, is trivial.  Instead of counting the number of agents in each space-time bin in our preprocessing procedure, we would calculate transitions, or flow counts, of  the number of agents that transition from spatial cell $i$ to spatial cell $j$ at time $t$ and would fit a DGLM for every observed transition between spatial cells (including self-transitions). This would greatly increase the number of DGLMs to fit, which is why we did not consider this setting here, but otherwise would require no updates to our modeling approach.  We would expect similar modeling results to those shown below, with the observations of occupancy counts simply being updated to flow counts.  In addition to increasing the number of DGLMs, modeling transitions instead of occupancy counts would increase the level of sparsity in each observed transition; however, DCMMs could still be applied to capture varying levels of sparsity in these transition counts.

An advantage of DGLMs for modeling human mobility data is the ease of incorporating external, dynamic, information into the DGLMs directly.  Here, we only use a trend term and daily seasonality for our synthetic data results, though we could easily incorporate holiday effects, point of interest information, etc. into the models.  We also only tune the DGLM discount factor hyper-parameters over a limited grid to speed-up computation, but could increase the degree of hyper-parameter tuning to further improve upon forecast accuracy results presented below. 

Finally, our approach requires a choice of temporal and spatial resolutions.  There is a trade-off between finer resolutions and the degree of sparsity for each spatial cell that is modeled (Figure~\ref{fig:sparsity}); while DCMMs can accurately model and forecast count time series with a variety of levels of sparsity, the more sparse the time series becomes, the more challenging it is to accurately predict where the non-zero observations will occur \cite{YanchenkoEtAl2021}.  To address this challenge, we could model over multiple spatial resolutions, so that the spatial cells were not uniform in area but were variably sized based on a base level of occupancy counts; this would only require changes to our preprocessing, but not our modeling approach.  Alternatively, we could form multi-scale DGLMs over multiple spatial and/or temporal resolutions \cite{doi:10.1080/01621459.2015.1129968, BerryWest2018DCMM, YanchenkoEtAl2021} to share information at more aggregate resolutions with the finer-grained resolutions; this information could be added as an additional covariate in our DGLMs with no further updates to the modeling approach.  Similarly, we could incorporate spatial dependence between cells by using the forecasted occupancy count of one cell as a covariate in the DGLM for another spatial cell.


\subsection{Anomaly Detection}
The focus of this work is on accurate probabilistic modeling and forecasting of occupancy counts, however our approach supports several additional downstream applications, including anomaly detection.  The specific manner in which to use the DGLMs for anomaly detection depends on the types of anomalies that need to be detected; modeling occupancy counts with DGLMs as presented here supports anomaly detection over aggregate changes in agent behavior, which is reflected in higher or lower occupancy counts than expected in a spatial cell.  For example, in a transportation traffic monitoring application, higher occupancy counts than expected could represent a traffic jam, while lower counts than expected could represent a road closure.  Anomalous occupancy counts can be detected by comparing  the quantile or likelihood of the observed occupancy count at time $t$ under the one-step ahead forecast distribution to the ``expected'' count at time $t$. This ``expected'' value could be a multi-step ahead forecast from further back in time or from another DGLM that does not adapt as quickly to current data.  DGLMs are probabilisitic models, which allows for uncertainty estimates to propagate into the anomaly detection, too.

\section{Results}
\subsection{Forecast Metrics and Examples}
Across a variety of spatial resolutions and forecast metrics, DGLMs accurately model the observed occupancy counts and the uncertainty estimates of the forecast distributions are well-calibrated (Table~\ref{tab:1step-forecasts}).  The RMSE and MAE forecast metrics are on the same scale as the data; on average, the 1-step ahead forecasts across spatial cells are only a handful of occupancy counts off from the observations, and the average percent error is also  low for the finer resolution spatial cells.  The coverage of the DGLMs at 90\% tends to be > 90\%, indicating that the uncertainty estimates are slightly wider than they need to be; in practice, over-coverage is often preferred to under-coverage.  Additional tuning of the discount factors, specifically the discount factor for the random effects in the Poisson DGLM and DCMM \cite{BerryWest2018DCMM},  could improve the calibration of the uncertainty estimates.  However, with a single modeling framework, we are able to model a variety of different occupancy counts, demonstrating the flexibility of our approach.

\begin{figure}[h]
  \centering
  \includegraphics[width=0.8\linewidth]{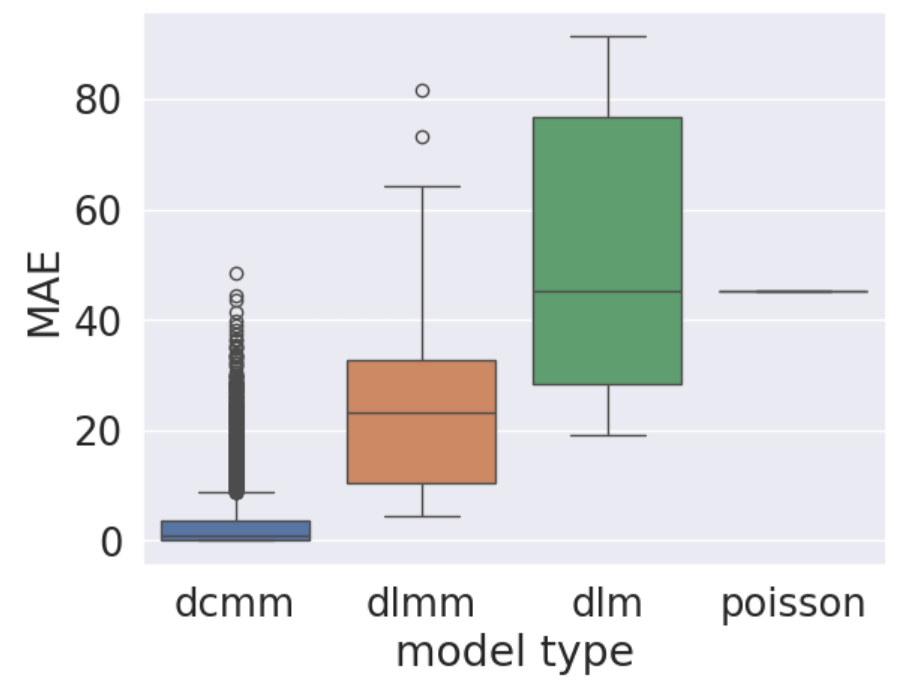}
  \caption{Mean absolute error (across spatial cells) by model type for S2 cell level 14. DLMs are used to model higher occupancy count cells and can have larger forecast errors than the DCMMs, used for low and sparse occupancy count cells.}
   \label{fig:metrics-boxplot}
\end{figure}

The magnitude of the observed occupancy counts does impact the forecast accuracy of the models (Figure~\ref{fig:metrics-boxplot}); the DLMs are used to model spatial cells with higher occupancy counts and the forecast medians have more potential to differ from the observed occupancy counts.  On the other hand, the DCMMs are used for lower count spatial cells, leading to lower forecast metric values.  The finer the spatial resolution, the lower the occupancy counts for each spatial cell overall, which is why the point forecast metrics in Table~\ref{tab:1step-forecasts} improve with finer spatial resolutions.  However,  at finer spatial resolutions, sparsity also increases.  When the degree of sparsity for each spatial cell is particularly high, e.g. > 70\% of counts for a single spatial cell are zero, it is challenging to predict when the non-zero occupancy counts will occur.  Modeling could be improved in these settings with additional covariates; for example,  point-of-interest information could aid in predicting a zero vs. non-zero count for these sparse cells.  The level of noise in the data is also a consideration; as the spatial resolution becomes finer, it is more likely that sensor noise will cause occupancy counts to be split across spatial cell boundaries.  This can be addressed in several ways; for example, smoothing the raw trajectory data, only encoding one observation per agent per time bin, or considering multi-scale approaches over multiple spatial resolutions \cite{doi:10.1080/01621459.2015.1129968}.  In practice,  the specific application goals will guide the choice of spatial and temporal resolutions for modeling the occupancy counts.

\begin{table*}
  \caption{Forecast accuracy metrics across spatial cells for 1-step ahead forecasts.  Mean and standard deviation across all cells is shown for the forecast median. RMSE: root-mean squared error, MAE: mean absolute error, ZAPE: zero-adjusted percent error.  Lower is better for RMSE, MAE, and ZAPE and closer to the nominal coverage (e.g. 80\%, 90\% or 95\%) indicates better calibrated coverage.}
  \label{tab:1step-forecasts}
  \begin{tabular}{cccccccc}
    \toprule
    S2 Cell Resolution& RMSE & MAE & ZAPE & 80\% Coverage &90\% Coverage & 95\% Coverage & \# of Models\\
    \midrule
    10 & $65.62 \pm 87.24$ & $ 36.17\pm 57.73$ & $63.53\pm 137.85$ & $92.40\pm 4.48$ &  $94.87\pm 3.35$ & $96.04\pm 2.98$ & 103 \\
    14 & $9.18\pm 14.68$ & $3.81\pm 7.15$ & $22.87\pm 40.51$ & $94.37 \pm 5.99$ &  $95.87 \pm 4.62$ & $96.62 \pm 4.16$ & 3938\\
    17 & $1.10 \pm 2.15$ & $0.37 \pm 1.03$ & $8.49 \pm 8.99$  & $97.32 \pm 2.46$ &  $98.38 \pm 1.72$ &  $98.97\pm 1.35$ & 42038 \\
  \bottomrule
\end{tabular}
\end{table*}

\begin{figure}[h]
  \centering
  \includegraphics[width=0.8\linewidth]{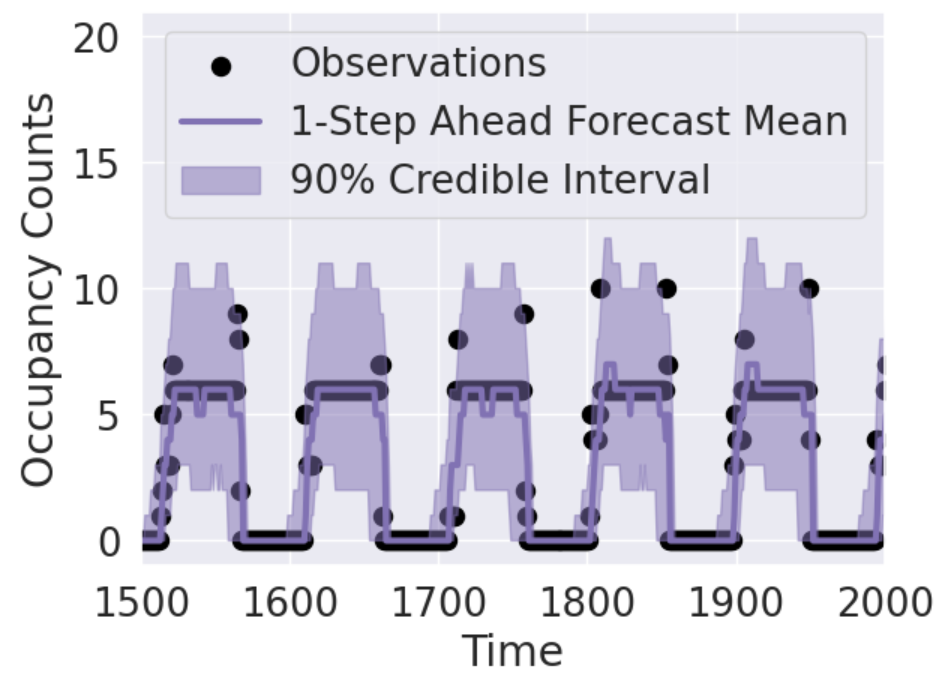}
  \caption{One-step ahead forecast mean and credible intervals for a DCMM fit to a spatial cell with low, sparse occupancy counts at S2 cell level 14.}
   \label{fig:DCMM2}
\end{figure}

DGLMs accurately model and forecast a variety of levels of occupancy counts, including high counts fit with a DLM, moderate counts with sparsity fit with a DLMM and sparse, lower levels of counts fit with a DCMM (Figure~\ref{fig:DCMM2}).  Across these example spatial cells, the forecast means closely track the observed occupancy counts and the 90\% credible intervals are well calibrated, as seen across all spatial cells in Table~\ref{tab:1step-forecasts}. The flexibility of the DCMMs and DLMMs, in particular, to model spatial cells with a high-level of sparsity (or 0 occupancy counts), is critical, as across spatial resolutions, we observe many spatial cells with some degree of sparsity, varying from minimal sparsity to more than 80\% sparsity (Figure~\ref{fig:sparsity}).  In our synthetic data, this sparsity occurs as a result of zero counts in a given spatial cell, but the DGLMs also easily handle missing data by not updating for timesteps that have no observations.  The majority of spatial cells in our synthetic data exhibit strong daily seasonality, which is explicitly incorporated into each of the DGLMs and could be expanded to include seasonality at additional resolutions, like weekly or monthly.

\begin{figure*}[h]
  \centering
  \includegraphics[width=0.8\textwidth]{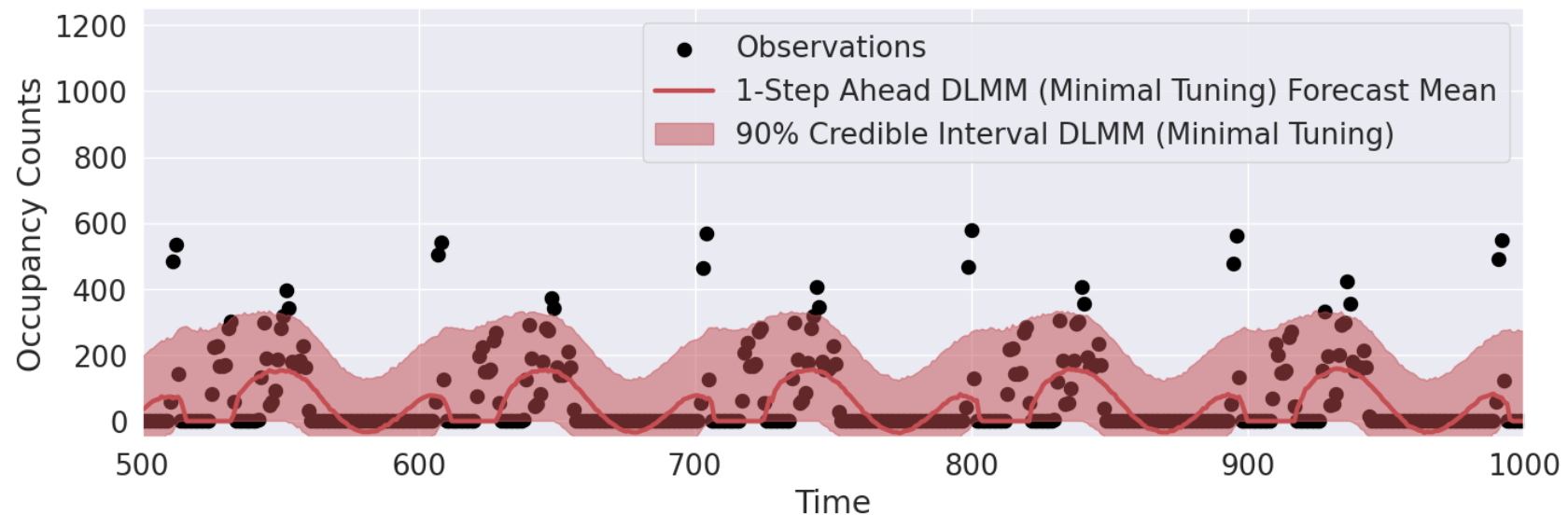}
  \caption{One-step ahead forecast mean and credible intervals for a S2 level 10 cell modeled with a DLMM. This cell is challenging to model, due to the high number of zero counts and the high values of non-zero occupancy counts.  For the base DLMM, where the only discount factor tuned is the trend term over a grid search of $[0.96, 0.97]$,  the RMSE for the 1-step ahead forecast mean is 109.75, the MAE is 70.91, the ZAPE is 984.40, and the 90\% empirical coverage is 95.31\%. }
   \label{fig:dlmm-example}
\end{figure*}

\begin{figure*}[h]
  \centering
  \includegraphics[width=0.8\textwidth]{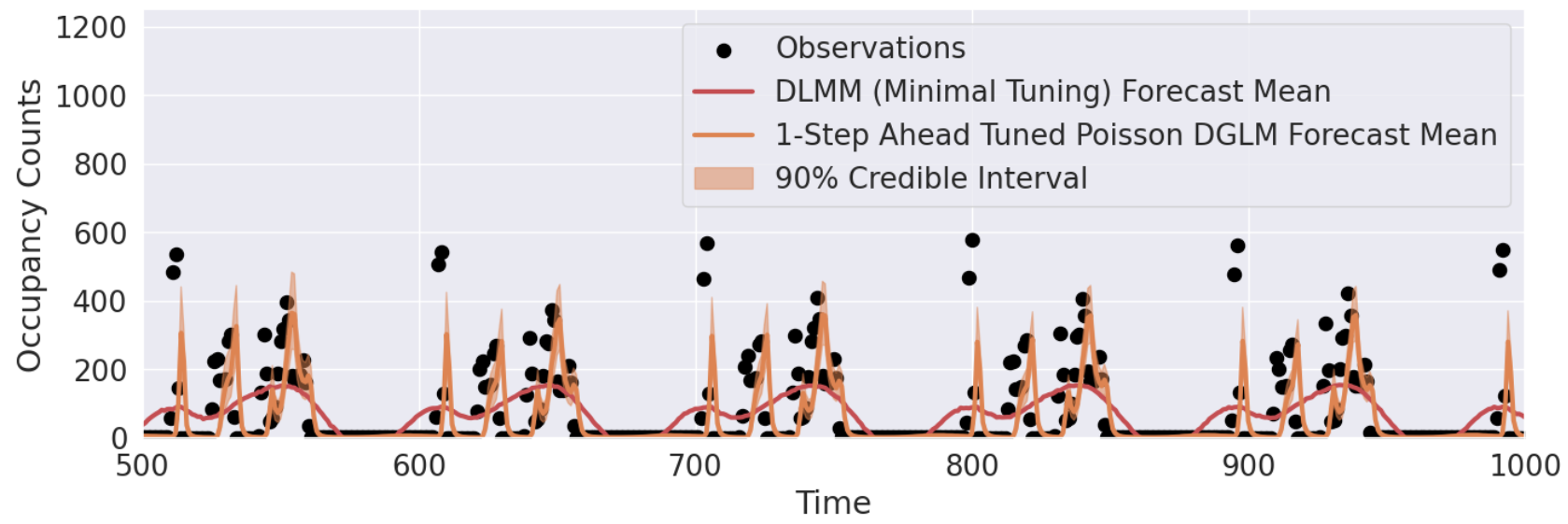}
  \caption{One-step ahead forecast mean and credible intervals for the same S2 level 10 cell in Figure~\ref{fig:dlmm-example}, this time modeled with a Poisson DGLM. The trend discount factor is tuned over $[0.9, 0.93, 0.95, 0.96, 0.97, 0.98, 0.99, 0.999, 1]$ and the random effect discount factor over $[0.6, 0.7, 0.8, 0.85, 0.9, 0.95, 0.99, 1]$.  While the point forecast metrics improve slightly with this Poisson DGLM, the uncertainty estimates are much worse.  The RMSE for the 1-step ahead forecast mean is 117.93, the MAE is 62.12, the ZAPE is 221.91, and the 90\% empirical coverage is 50.0\%.}
   \label{fig:dlmm-example-poisson}
\end{figure*}

\begin{figure*}[h]
  \centering
  \includegraphics[width=0.8\textwidth]{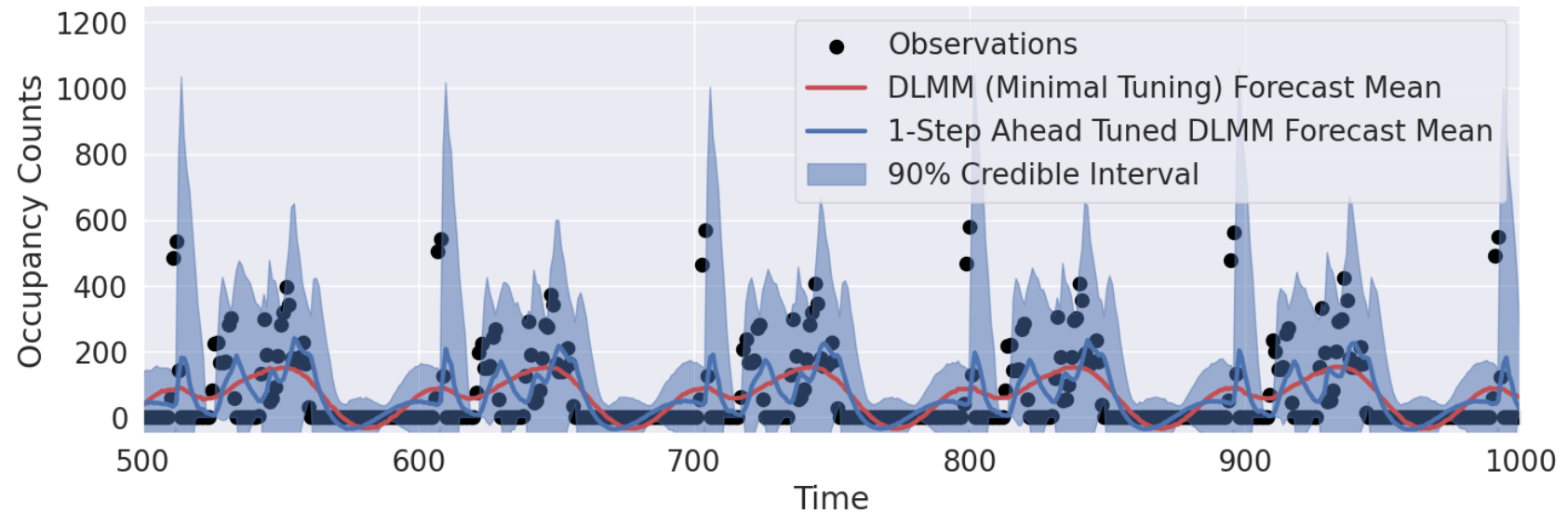}
  \caption{One-step ahead forecast mean and credible intervals for the same S2 level 10 cell in Figure~\ref{fig:dlmm-example} for a tuned DLMM. The trend discount factor is tuned over $[0.9, 0.93, 0.95, 0.96, 0.97, 0.98, 0.99, 0.999, 1]$ and the stochastic variance discount factor over $[0.6, 0.7, 0.8, 0.85, 0.9, 0.95, 0.99, 1]$.  The uncertainty estimates are slightly better in this case, compared to the coverage in Figure~\ref{fig:dlmm-example} and the point forecast metrics are similar.  The RMSE for the 1-step ahead forecast mean is 114.42, the MAE is 77.77, the ZAPE is 956.41, and the 90\% empirical coverage is 93.0\%}
   \label{fig:dlmm-example-tuned}
\end{figure*}

Spatial cells with high levels of counts and high levels of sparsity are in general challenging to model; these cells contribute to the larger forecast errors for S2 level 10 spatial cells in Table~\ref{tab:1step-forecasts}.  An example spatial cell is shown in Figure~\ref{fig:dlmm-example}. This spatial cell exhibits relatively high, non-zero counts (82.2 on average) and a large degree of sparsity; 19\% of time steps have zero occupancy counts.  A DLMM is automatically selected to model this spatial cell, and the model has relatively high forecast error.  This cell is challenging to model because of the sparsity and because the non-zero counts are not normally distributed.  Alternatively, we could use a Poisson DGLM to model this spatial cell (Figure~\ref{fig:dlmm-example-poisson}); while this improves the MAE and ZAPE metrics, coverage is significantly worse.  Even allowing for over-dispersion, the Poisson distribution has only one parameter for the mean and variance.  In these high count settings, the Poisson DGLM significantly under-estimates the uncertainty in the occupancy counts.  Alternatively, we can tune the trend and stochastic variance discount factors over a larger grid of values to improve the forecast accuracy (Figure~\ref{fig:dlmm-example-tuned}).  The optimal discount factors are much lower than the values selected in Figure~\ref{fig:dlmm-example}; 0.9 for the discount factor for the trend and 0.6 for the stochastic variance in  Figure~\ref{fig:dlmm-example-tuned} compared to 0.96 for the discount factor for the trend and 0.9 for the stochastic variance in  Figure~\ref{fig:dlmm-example}.  Lower values for the stochastic variance increase how much over-dispersion is expected in the counts, leading to the improved uncertainty estimates.  Additionally, the lower discount factor for the trend allows the model to adapt more quickly to the incoming data, leading to a slight improvement in the forecast accuracy.  Finally, the forecast accuracy for these cells could likely be further improved by additional preprocessing to make the normal distribution assumption for the non-zero counts more valid, or considering different distributions than a mixture of a Bernoulli DGLM and a normal DLM via the DLMM.  However, our results here demonstrate the flexibility and feasibility of our approach to a variety of choices of spatial and temporal resolutions, and across different DGLM model types.

\subsection{Multi-Step Ahead Forecasting}

Multi-step ahead forecasting is also straightforward with DGLMs, and the DGLM point forecasts remain accurate across longer-range forecast horizons (Table~\ref{tab:kstep-forecasts}).  Our temporal resolution is 15 minutes for the mobility data, and we compare results for 1 timestep ahead to forecast results for 1 hour (4 steps ahead), 6 hours (24 steps ahead) and 24 hours (96 steps ahead).  Forecast accuracy across metrics is comparable  at all of these forecast horizons, demonstrating the ability of DGLMs to be used for long-range forecasting for human mobility.  Results below are for marginal, rather than path, forecasts \cite{West1997, BerryWest2018DCMM}.  The coverage is lower at multi-step ahead forecast horizons because we do not propagate the state vector uncertainty (via the discount factors) at longer than 1-step ahead forecast horizons in these results, though that additional uncertainty could easily be incorporated to the multi-step ahead forecasts.  We could alternatively use path forecasts to improve the uncertainty estimates, at the expense of longer computation times than the marginal forecasts shown here.   The long-range forecast accuracy benefits from incorporating the strong daily seasonality explicitly into the DGLMs, which is very straightforward to do with these models.  Other relevant external information could also be easily incorporated to further improve forecast accuracy as needed.

\begin{table*}
  \caption{Forecast accuracy metrics across S2 level 14 spatial cells for multi-step ahead forecasts.  Table format as in Table~\ref{tab:1step-forecasts}.}
  \label{tab:kstep-forecasts}
  \begin{tabular}{cccccccc}
    \toprule
    Forecast Steps Ahead  & Time Ahead  & RMSE & MAE & ZAPE & 80\% Coverage & 90\% Coverage & 95\% Coverage  \\
    \midrule
     $1$ & 15 min & $9.18\pm 14.68$ & $3.81\pm 7.15$ & $22.87\pm 40.51$ & $94.37 \pm 5.99$ &  $95.87 \pm 4.62$ & $96.62 \pm 4.16$ \\
    $4$ &  1 hour & $ 11.20\pm 18.45 $ & $4.76 \pm 8.88$ & $23.02 \pm 38.74$ & $73.22 \pm 4.37$ & $74.39\pm 3.58$ & $75.04\pm 3.21$ \\
    $24$ & 6 hours & $9.06 \pm 15.00$ & $3.68 \pm 7.35$ & $20.79 \pm 38.44$ & $73.26 \pm 3.51$ & $74.17\pm 2.98$ & $74.72\pm 2.67$   \\
    $96$ & 24 hours & $9.35 \pm 14.87$ & $3.91 \pm 7.31$ & $23.27 \pm 38.71$ & $69.46 \pm 4.56$ & $70.91\pm 3.40$ & $71.49 \pm 3.08$  \\
  \bottomrule
\end{tabular}
\end{table*}

\begin{figure}[h]
  \centering
  \includegraphics[width=0.8\linewidth]{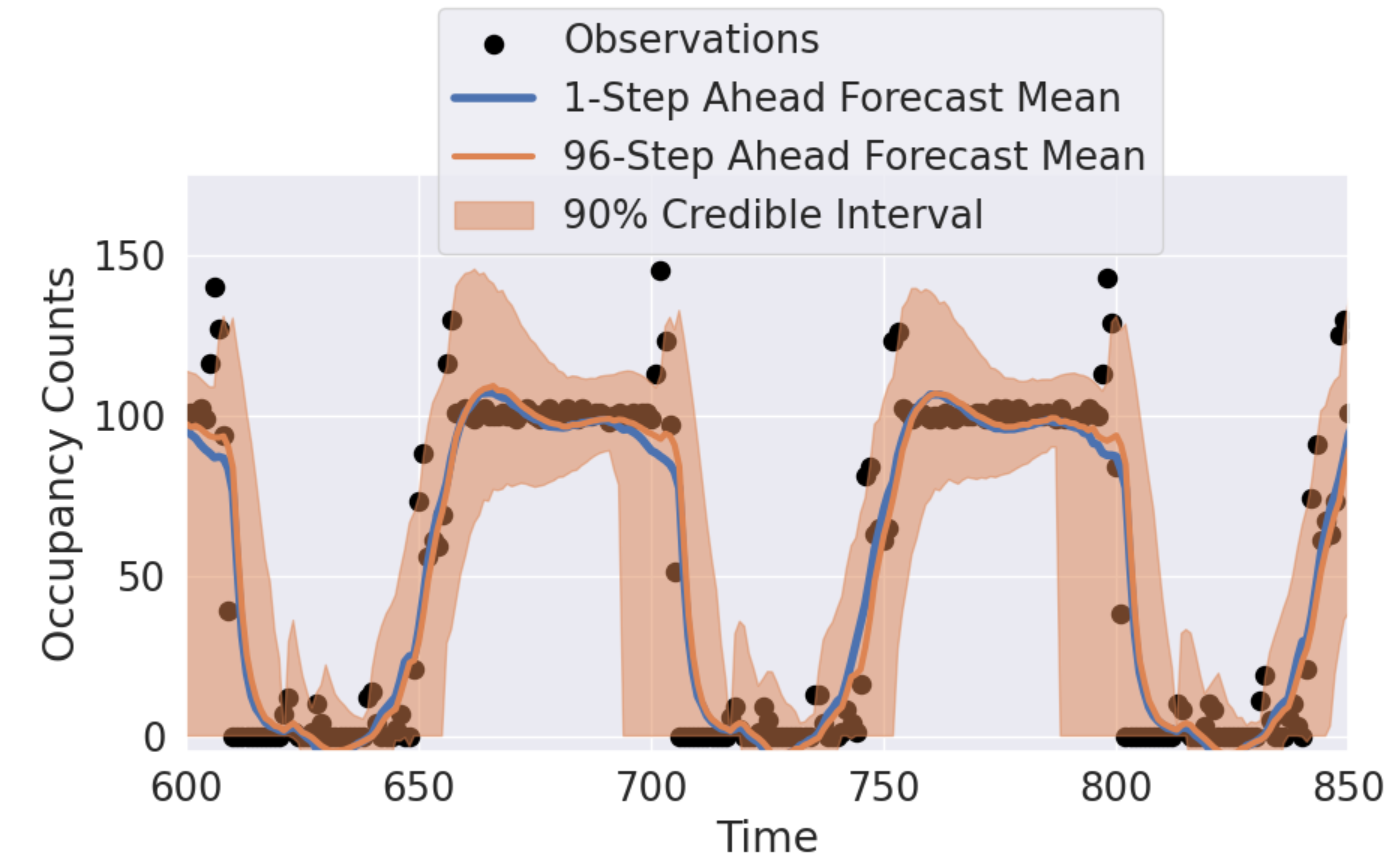}
  \caption{Twenty four-hour ahead (96-step ahead) forecast mean and credible intervals, compared to the 1-step ahead forecast mean, for a DLMM fit to a spatial cell at S2 level 14.}
   \label{fig:multistep-k96-v2}
\end{figure}

Figure~\ref{fig:multistep-k96-v2} show twenty-four hour ahead forecast results (96-steps ahead) for a DLMM fit to a relatively sparsely occupied spatial cell at S2 level 14.  The forecasts at this long forecast-horizon are accurate, do not degrade much from the 1-step ahead forecasts, and still have well calibrated forecast uncertainty.  These results demonstrate the flexibility of the mixture models (DLMMs and DCMMs), in particular, to accurately model very sparse occupancy counts and the ability of DGLMs to accurately forecast over long time horizons; important qualities when using these models for a variety of downstream applications.

\subsection{Robustness}
The synthetic data used in the experiments above has very regular seasonality in the occupancy counts. However, our approach is flexible to both different types of seasonality and less regularity in the seasonality.  To demonstrate this feature of the DGLMs, we also include robustness results that focus on adding noise to the period of the daily seasonality in the occupancy counts.  For each spatial cell at S2 level 14, we randomly sample a slight offset independently for each day to shift the seasonal pattern slightly forward or slightly backward in time (we also include no shift as an option to randomly select).  This additional noise in the data is meant to represent real world data, where there may be regular seasonality, but the period of the seasonality is not as regular as the synthetic data we work with here; each day the occupancy count peak might shift sooner or later in the day, and this pattern is random.  

We add this random noise in all spatial cells at S2 level 14 for each day of data, over four possible levels of offset: up to 0.5 hours, up to 1 hour, up to 2 hours, and up to 3 hours sooner or later than expected.  For all settings, no offset is also a possible noise setting.  We then fit the DGLMs as described above to all spatial cells and compare to a baseline model that just uses the occupancy count $t$ - 24 hours  ago to predict the occupancy count at time $t$.  The DGLMs still have daily seasonality in the model and we do not change any experimental settings except to retune the trend discount factor for the noisy data.  Except for retuning and fitting, the modeling setup is identical to the results above without seasonal noise.  

\begin{figure}[h]
  \centering
  \includegraphics[width=0.8\linewidth]{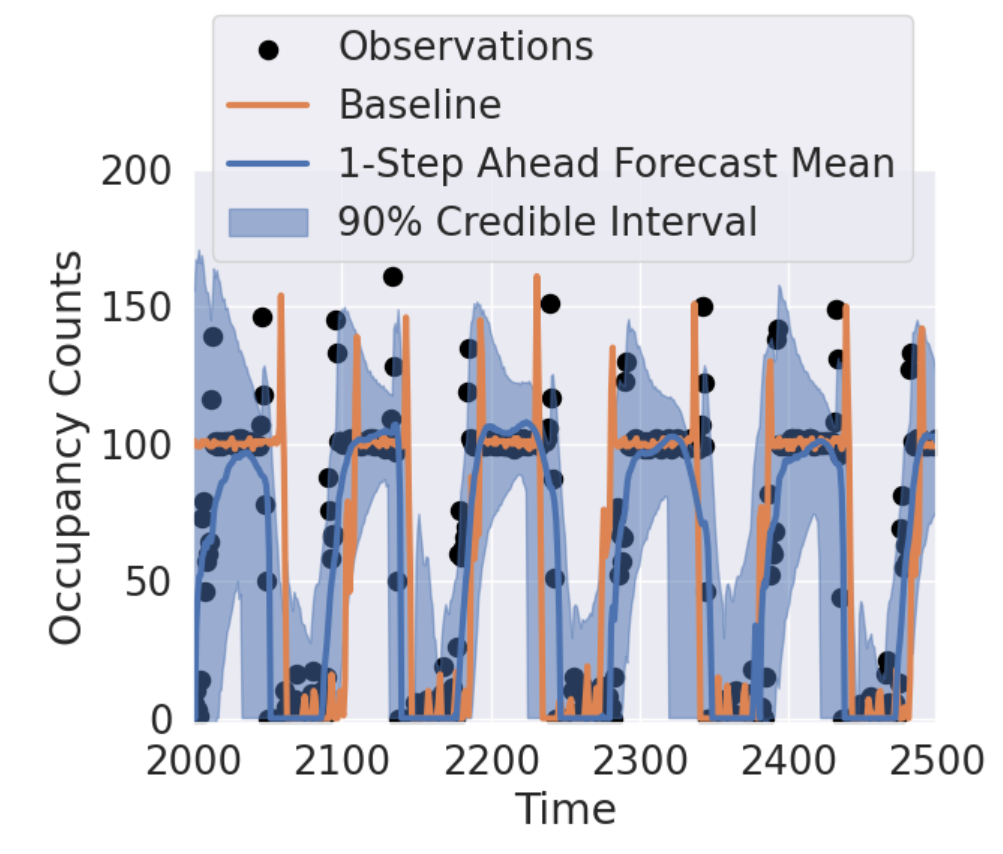}
  \caption{One-step ahead forecast mean and credible intervals, compared to the seasonal baseline, for a DLMM fit to a spatial cell at S2 level 14 with seasonal offset up to 3 hours.  The DLMM is better able to adapt to slightly varying seasonal periods.}
   \label{fig:noise-dlmm}
\end{figure}

In addition to providing uncertainty estimates, the DGLMs are able to flexibly adapt to noise in the seasonal pattern, across different levels of noise offset, while the simple baseline models are not.  Forecast metrics for these experiments are shown in Table~\ref{tab:noise} and example forecasts are shown in Figure~\ref{fig:noise-dlmm} and Figure~\ref{fig:noise-dcmm-2}.  DGLM forecast accuracy remains the same throughout these offsets, while the baseline forecast accuracy degrades with larger offsets. These results demonstrate the utility of our approach on realistic data, that is expected to have more variation in seasonality.   

\begin{table}
  \caption{Forecast accuracy metrics across S2 level 14 spatial cells with noise added to the daily seasonal period. For example, an offset of 1 hour means that the possible daily seasonal periods were 23, 23.5, 24, 24.5, and 25 hours; this period was randomly chosen for each day for each spatial cell.  Table format as in Table~\ref{tab:1step-forecasts}.}
  \label{tab:noise}
  \begin{tabular}{cccccccc}
    \toprule
    Offset  & Model  & RMSE & MAE & ZAPE   \\
    (hrs.)  &   &  &  &    \\
    \midrule
     $0.5$ & DGLM &  $\bm{9.26\pm 14.91}$ & $3.86\pm 7.32$ & $\bm{23.23\pm 42.21}$  \\
      & Baseline &  $10.07\pm 15.79$ & $3.57\pm 6.00$ & $26.04\pm 29.80$  \\\midrule
    $1$  & DGLM &  $\bm{9.38\pm 15.33} $ & $\bm{3.94 \pm 7.62}$ & $\bm{23.39 \pm 41.89}$  \\
     & Baseline &  $11.51\pm 18.92$ & $4.45\pm 8.10$ & $28.26\pm 34.39$  \\\midrule
    $2$  & DGLM & $\bm{9.74 \pm 16.84}$ & $\bm{4.08 \pm 8.36}$ & $\bm{23.19 \pm 47.92}$    \\
 & Baseline &  $12.65\pm 21.24$ & $5.25\pm 9.89$ & $29.35\pm 39.00$  \\\midrule
    $3$  & DGLM  & $\bm{9.93 \pm 17.31}$ & $\bm{4.15 \pm 8.89}$ & $\bm{22.48 \pm 47.39}$   \\
     & Baseline &  $13.55\pm 22.99$ & $5.81\pm 11.34$ & $31.09\pm 54.95$  \\
  \bottomrule
\end{tabular}
\end{table}

\subsection{Scaling and Implementation}

Results shown above were for  $\sim$11k agents over one month of data.  This simulated data was 9.77 GB, compressed and at rest and results were run on a machine with 48 CPU cores.  Computation and memory requirements for the    $\sim$11k agents are shown in Table~\ref{tab:scaling}; our preprocessing procedure significantly reduces the size of the input data to model. Additionally, preprocessing time is approximately constant across spatial resolutions, as expected, and modeling time increases linearly with the number of spatial cells to model; we fit and forecast DGLMs across $\sim$ 4k spatial cells in just over 1 hour.

We have also applied our approach at a much larger scale and are still able to preprocess and model the data efficiently.  With our approach, the preprocessing time depends on the number of raw observations and not the choice of spatial or temporal resolutions for the modeling. Our preprocessing also significantly reduces the size of the data to model.  We applied this approach to additional synthetic data with \textbf{484 million} observations 
which was 472 GB of data compressed and at rest.  After transforming this data to occupancy counts with a spatial resolution of S2 cell level 17 and temporal resolution of 1 hour, the data size was reduced to \textbf{0.582 GB} compressed.  Preprocessing the 484 million observations took 10 hours on a 384-core cluster. We then fit  397352 DGLMs on the same size cluster in 10 hours; each DGLM updated and forecasted in an average of 2.5 seconds for 672 timesteps, or approximately 4 ms per DGLM per timestep to update and forecast.  

\begin{table*}
  \caption{Computation time and memory requirements for the results shown on $\sim$11k agents for 28 days; the input data was 9.77 GB compressed and at rest. Preprocessing significantly reduces the size of the data to model and preprocessing time is the same across spatial resolutions; modeling time scales linearly with the number of spatial cells/models.}
  \label{tab:scaling}
  \begin{tabular}{cccccc}
    \toprule
    S2 Cell Resolution& Preprocessing  & Modeling  & Total  &Preprocessing  & \# of Models\\
    &  Time (Min.) &  Time (Min.) &  Time (Min.) & Output (MB) &  \\
    \midrule
    10 & 15.17 & 18.84 & 34.01 & 0.32& 103 \\
    14 & 16.45 & 66.48 & 82.94 & 7.0 & 3938 \\
    17 & 16.49 & 526.92 & 543.41 & 49.0 & 42038 \\
  \bottomrule
\end{tabular}
\end{table*}

Our approach was implemented in PySpark \cite{PySpark}, with Apache Sedona \cite{Sedona, Yu:2019vt} to encode the data into S2 cells and PyBats \cite{PyBats} to fit the DGLMs.  Further efficiencies are possible with additional tuning of the Spark cluster and improvements to the implementation of the DGLMs.  However, our results show the feasibility of our approach for very large scale datasets both in terms of the preprocessing and the modeling; our preprocessing scales linearly with the number of observations and the modeling scales linearly with the number of space-time bins.  Our approach is very well-suited to a distributed computing environment, as each model only operates on a small portion of the data and is independent of all other models.

\section{Summary and Future Work}

In this work, we present a scalable and flexible approach to modeling and forecasting human mobility data.  Our approach encodes the spatiotemporal data as occupancy counts in space-time bins and models these counts with DGLMs.  DGLMs provide efficient inference, probabilistic modeling and forecasting, flexibility to a range of observed occupancy count levels (including very sparse occupancy counts and over-dispersed counts), and can easily incorporate external information, such as seasonality.  We present forecast results on synthetic human mobility data for $\sim$11k agents; DGLMs maintain high forecast accuracy for a range of spatial resolutions and for long forecast horizons.    Finally, we show scaling results for much larger synthetic data consisting of hundreds of millions of observations to demonstrate the scalability of our approach.

Our approach is flexible enough to support a variety of downstream applications that require accurate, probabilistic forecasting and modeling.  For example, the forecast results could support traffic or transportation management, location demand forecasting, and urban planning.  Additionally, the flexibility of our approach makes it applicable to other types of spatiotemporal data beyond human mobility,  for example, our approach could aid in modeling animal movements at scale.  

There are several directions for future work and extensions of our approach.  One main direction, which would be largely application driven, is to incorporate additional information into the DGLMs.  For example, the mobility data may include holiday effects that need to be accounted for, different patterns of seasonality, or we may wish to incorporate dynamic covariates for each spatial cell, related to point of interest information, for example.   We could easily extend our approach to modeling flows of agents between spatial cells to better support movement or transportation related applications. 

On the modeling side, a main direction for future work involves improving the modeling of very sparse spatial cells, especially cells that are both sparse and have large non-zero counts, as in Figure~\ref{fig:dlmm-example}.  Although DCMMs and DLMMs can accurately model a variety of sparse count levels, extremely sparse counts are challenging to model and forecast well; sparse time series modeling is a challenge in many different settings \cite{YanchenkoEtAl2021}.  One solution would be to model at multiple resolutions or uneven spatial resolutions, to prevent any spatial cells from being too sparse.  Another solution would be to share information over multiple spatial resolutions (and/or temporal resolutions) with a multi-scale approach \cite{doi:10.1080/01621459.2015.1129968, BerryWest2018DCMM, YanchenkoEtAl2021}.   

Another area for future work is faster tuning of the discount factors than a grid search-based approach.  For the majority of spatial cells, more fine-grained tuning of all discount factors improves forecast accuracy.  However, with a grid search, we have to fit a DGLM for each combination of discount values to tune; it can be expensive to tune over too many unique values.  Using a coarse approximation to this optimization problem could improve the forecast accuracy while also improving the scalability of our approach.  


\begin{acks}
Supported by the Intelligence Advanced Research Projects Activity (IARPA) via Department of Interior/ Interior Business Center (DOI/IBC) contract number 140D0423C0046. The U.S. Government is authorized to reproduce and distribute reprints for Governmental purposes notwithstanding any copyright annotation thereon. Disclaimer: The views and conclusions contained herein are those of the authors and should not be interpreted as necessarily representing the official policies or endorsements, either expressed or implied, of IARPA, DOI/IBC, or the U.S. Government.  
\end{acks}

\bibliographystyle{ACM-Reference-Format}
\bibliography{abbrev}

\appendix

\section{Implementation Details}

\subsection{Summary of Overall Approach}\label{sec:details}

\textbf{Input:} GPS trajectory data with ID, latitude, longitude, and timestamp observations
\begin{enumerate}
\item \textbf{Data Preprocessing:} encode trajectory data into occupancy counts following Figure~\ref{fig:preprocessing}. Result: occupancy counts $y_t^i$ for every spatial cell $i\in {1, \ldots, I}$ and time bin $t\in {1, \ldots, T}$.
\item \textbf{Initialize DGLMs:} For all spatial cells $i\in {1, \ldots, I}$:
\begin{enumerate}
\item Select the first $\tilde{t}$ timesteps for initialization (in results above, $\tilde{t} =$ 288, or initialization over the first 72 hours with 15 minute temporal resolution) 
\item Select the observation distribution based on the mean of the first $\tilde{t}$ occupancy counts, $\bar{y}^i_{1:\tilde{t}} = \frac{1}{\tilde{t}}\sum_{j=1}^{\tilde{t}}y^i_j$, and the sparsity of these occupancy counts $s^i_{1:\tilde{t}} = \frac{1}{\tilde{t}}\sum_{j=1}^{\tilde{t}}\bm{1}(y^i_j = 0)$:
\begin{itemize}
\item If $\bar{y}^i_{1:\tilde{t}} > 50$ and $s^i_{1:\tilde{t}} < 0.15$, select a DLM.
\item If $\bar{y}^i_{1:\tilde{t}} \leq 50$ and $s^i_{1:\tilde{t}} < 0.15$, select a Poisson DGLM.
\item If $\bar{y}^i_{1:\tilde{t}} > 50$ and $s^i_{1:\tilde{t}} \geq 0.15$, select a DLMM.
\item If $\bar{y}^i_{1:\tilde{t}} \leq 50$ and $s^i_{1:\tilde{t}} \geq 0.15$, select a DCMM.
\end{itemize}
\item Set the covariate vector $\bm{F}_t$ and state matrix $\bm{G}_t$ based on the trend, covariate, and seasonal terms in the model; these values are known ahead of time and $\bm{F}_t$ and $\bm{G}_t$ are not learned parameters.  In this work, we use a linear trend term, no covariates, and daily seasonality with the first two harmonics (with 15 minute time bins, the period of the seasonality, $p$, is 96), $\bm{F}_t = \bm{F}$, $\bm{G}_t = \bm{G}$ and: 
$$\bm{F} = \begin{bmatrix}
1 & 1 & 0 & 1 & 0 \\
\end{bmatrix}',$$
$$\bm{G} = \begin{bmatrix}
1 & 0 & 0 & 0 & 0\\
0  & \mbox{cos}\left(2\pi/p\right) & \mbox{sin}\left(2\pi/p\right) & 0 & 0 \\
0 & -\mbox{sin}\left(2\pi/p\right) & \mbox{cos}\left(2\pi/p\right) & 0 & 0 \\
0 & 0 & 0 & \mbox{cos}\left(4\pi/p\right) & \mbox{sin}\left(4\pi/p\right) \\
0 & 0 & 0 & -\mbox{sin}\left(4\pi/p\right) & \mbox{cos}\left(4\pi/p\right) \\
\end{bmatrix},$$
\item Set and/or tune the discount factors. To tune the discount factors:
\begin{enumerate}
\item Select a grid of possible values for the discount factor(s) to be tuned.
\item For each value in the grid, initialize the appropriate DGLM with that discount factor value and update the model over all timesteps $t\in {1, \ldots, \tilde{t}}$. Save the one-step ahead forecasts and calculate the MAE.
 \item Select the discount factor over the grid resulting in the lowest MAE.
 \item We set the seasonality discount factor to be 0.994, the random effect discount factor or the stochastic variance discount factor to 0.9, and tune the trend discount factor over the grid of values $[0.96, 0.97, 1.0]$.
\end{enumerate} 
\item Result: an initialized DGLM for each spatial cell,  $\mathcal{M}^i$.
\end{enumerate}
\item \textbf{Modeling with DGLMs:} In parallel across all spatial cells $i$ and sequentially for $t\in {1, \ldots, T}$:
\begin{enumerate}
\item  Forecast $k$ steps ahead with $\mathcal{M}^i$ based on information up to time $t-1$.
\item Update model $\mathcal{M}^i$ using occupancy count $y_t^i$.
\end{enumerate}
\item \textbf{Analysis:} Analyze results and calculate metrics.
\end{enumerate}

Steps (2) and (3) are implemented in PyBats \cite{PyBats}.

\subsection{DGLM Sequential Learning}\label{subsec:dglm-learning}
DGLMs are Bayesian state space models (Equation~\ref{eq:dglm}), and learning proceeds sequentially through time.  Below are additional details for step (3b) in the implementation procedure in Section~\ref{sec:details} above for a Poisson DGLM.  Sequential analysis is analogous for the other DGLM forms, though the specific forms in steps (3) - (7) are unique to the specific exponential family form. Procedure following \cite{BerryWest2018DCMM, YanchenkoEtAl2021}; please see these references and \cite{West1997} for the general case.

For Poisson DGLM $\mathcal{M}^i$ at time $t$, $y^i_t\sim\mbox{Poisson}(\eta_t)$, $\lambda_t = \log \eta_t$:

\begin{enumerate}
\item State vector posterior from time $t-1$:\\ $(\bm{\theta}_{t-1} | y^i_{1:t-1}) \sim (\bm{m}_{t-1}, \bm{C}_{t-1})$; with $\mathbb{E}(\bm{\theta}_{t-1} | y^i_{1:t-1}) = \bm{m}_{t-1}$ and $\mathbb{V}(\bm{\theta}_{t-1} | y^i_{1:t-1}) = \bm{C}_{t-1}$.
\item State vector prior at time $t$: $(\bm{\theta}_{t} | y^i_{1:t-1}) \sim (\bm{a}_{t}, \bm{R}_{t})$; with $\bm{a}_t = \bm{G}_t\bm{m}_{t-1}$ and $\bm{R}_t = \bm{G}_t\bm{C}_{t-1}\bm{G}'_t + \bm{W}_t$.
\item Prior for the natural parameter at time $t$ (variational Bayes step): $(\eta_t | y^i_{1:t-1}) \sim \mbox{Gamma}(\alpha_t, \beta_t)$.
\item Evaluate the prior hyper-parameters $\alpha_t$ and $\beta_t$ such that:
$\mathbb{E}(\lambda_t | y^i_{1:t-1}) = f_t = \bm{F}_t'\bm{a}_t \mbox{ and } \mathbb{V}(\lambda_t | y^i_{1:t-1}) = q_t = \bm{F}_t'\bm{R}_t\bm{F}_t.$ For the Poisson DGLM, $f_t = \psi(\alpha_t) - \log\beta_t$, and $q_t = \psi'(\alpha_t)$, where $\psi$ is the digamma function and $\psi'$ is the first derivative of the digamma function. We can use numerical optimization, like Newton-Raphson, to solve for $\alpha_t$ and $\beta_t$.
\item Forecast $y^i_t$ one-step ahead: \\ $p(y^i_t |y^i_{1:t-1}) = \mbox{Negative Binomial}(\alpha_t, \beta_t/(1 + \beta_t))$.
\item Posterior for $\eta_t$ at time $t$: $p(\eta_t | y^i_{1:t}) = \mbox{Gamma}(\alpha_t + y_t^i, \beta_t + 1)$.
\item Map back to the linear predictor $\lambda_t = \log \eta_t$: posterior mean $g_t = \mathbb{E}(\lambda_t | y^i_{1:t}) = \psi(\alpha_t + y^i_t) - \log (\beta_t + 1)$ and posterior variance $p_t = \mathbb{V}(\lambda_t | y^i_{1:t}) = \psi'(\alpha_t + y^i_t)$.
\item State vector posterior at time $t$: $(\bm{\theta}_{t} | y^i_{1:t}) \sim (\bm{m}_{t}, \bm{C}_{t})$, with posterior mean $\bm{m}_t = \bm{a}_t + \bm{R}_t\bm{F}_t(g_t - f_t)/q_t$ and posterior variance $\bm{C}_t = \bm{R}_t - \bm{R}_t\bm{F}_t\bm{F}'_t\bm{R}_t' (1 - p_t/q_t)/q_t$.
\end{enumerate} 

The initial state vector mean is $\bm{m}_0 = \bm{0}$ and the initial state vector covariance is $\bm{C}_0 = \bm{I}$, the identity matrix. 

\subsection{Forecasting}

One-step ahead forecast distributions are calculated using step (5) in Section~\ref{subsec:dglm-learning} (and the analogous form for the other DGLMs).  The marginal multi-step ahead forecasts are calculated similarly.  In the multi-step ahead case, step (2) in    Section~\ref{subsec:dglm-learning} is updated by iterating the state vectors forward in time; for $k$-step ahead forecasts, step (2) becomes: $\bm{a}_{t+k} = \bm{G}^k\bm{m}_{t-1}$, $\bm{R}_{t+k} = \bm{G}^k\bm{C}_{t-1}(\bm{G}^k)'$, where $\bm{G}^k = \bm{G}_t\times\bm{G}_{t+1}\times\ldots\times\bm{G}_{t+k}$ is the matrix power.  $\bm{W}_t$ can optionally be added to $\bm{R}_t$ to continue discounting while forecasting.  Steps (3) - (5) then proceed as in   Section~\ref{subsec:dglm-learning} to result in forecasts for $k$-steps ahead.  

\section{Additional Forecast Example}

\begin{figure}[h]
  \centering
  \includegraphics[width=0.8\linewidth]{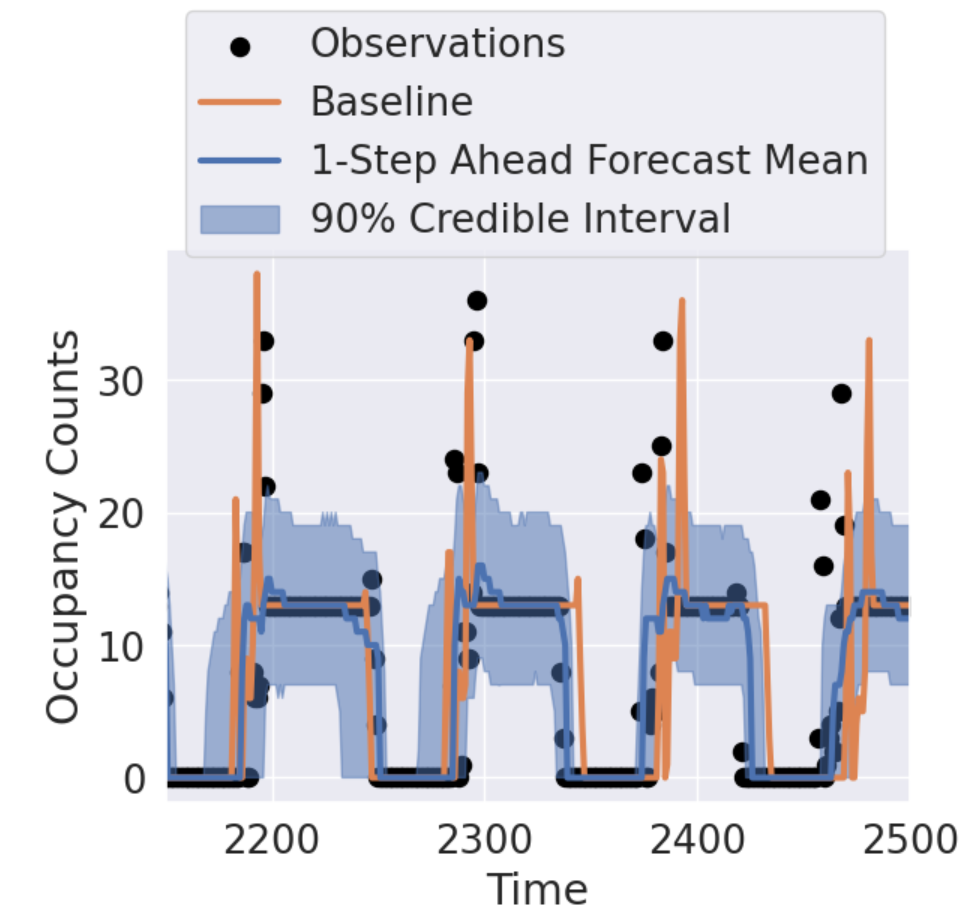}
  \caption{One-step ahead forecast mean and credible intervals, compared to the seasonal baseline, for a DCMM fit to a spatial cell at S2 level 14 with seasonal offset up to 3 hours.  The DCMM is better able to adapt to slightly varying seasonal periods.}
   \label{fig:noise-dcmm-2}
\end{figure}

We include an additional one-step ahead forecast example for a  DCMM compared to the seasonal baseline model (Figure~\ref{fig:noise-dcmm-2}).  The DGLM is able to adapt to the daily occupancy count spike, even when it occurs slightly less than or more than 24 hours from the previous spike; the seasonal baseline, which predicts the occupancy count at time $t$ as the occupancy count observed 24 hours ago is not able to adjust to noise in the seasonal period.  Additionally, the DGLM produces probabilistic forecasts, while the seasonal baseline just produces point forecasts.

\end{document}